# Energy translation and Proper-Time Eigenstates

**Zhi-Yong Wang[1], Cai-Dong Xiong[1], Ai-Lin Zhang[2]**

[1]School of Physical Electronics, University of Electronic Science and Technology of China, Chengdu, 610054, P.R. China

[2]Department of Physics, Shanghai University, ShangDa Road 99, Shanghai, 200444, P.R. China

E-mail: zywang@uestc.edu.cn

The usual quantum mechanics describes the mass eigenstates. To describe the proper-time eigenstates, a duality theory of the usual quantum mechanics was developed. The time interval is treated as an operator on an equal footing with the space interval, and the quantization of the space-time intervals between events is obtained. As a result, one can show that there exists a zero-point time interval.

## 1. INTRODUCTION

As we know, special relativity relates space with time via the Lorentz transformations and provides us the notion of a unified four-dimensional (4D) space-time. Quantum theory further relates generalized coordinates with its conjugate momentums by the Heisenberg uncertainty principle, and provides us the notion of a unified phase space, which is also exhibited by Fourier transformations, particularly, by the fractional Fourier transformations [1-2] and by the phase space representation of quantum mechanics [3]. Therefore, one could regard a coordinate space and its conjugate momentum space as two complementary aspects of the unified phase space. In fact, in a unified point of view, the general formulation of the law governing the motion of mechanical systems is the principle of least action, where the action always corresponds to some kind of measure of the phase space.

On the one hand, the usual quantum mechanics describes "particle" states or mass eigenstates [4-5]. Here a four-dimensional (4D) space-time coordinate $x^{\mu}=(t,\boldsymbol{x})$ corresponds to four independent variables, while only three components of a 4D momentum $p^{\mu}=(E,\boldsymbol{p})$ are independent because of the mass-shell relation $p^{\mu}p_{\mu}=m^{2}$ ($m$ denotes a mass), and the time coordinate $t$ plays the role of a parameter rather than that of a dynamical variable. We call the usual quantum mechanics position-space quantum mechanics.

On the other hand, for a given proper-time interval $\tau$, only three components of a 4D space-time interval $y^{\mu}=(y^{0},\boldsymbol{y})$ are independent because of the relation $y_{\mu}y^{\mu}=\tau^{2}$. If one introduces a wave function to describe the eigenstates of the proper-time interval $\tau$ (called the proper-time eigenstates), the mass would be completely indeterminate [6-7], therefore the proper-time eigenstates correspond to the off-mass-shell states rather than to the usual "particle" states, and can be called "event states". For the moment, contrary to the usual quantum mechanics, a 4D "momentum coordinate" $w^{\mu}=(w_{0},\boldsymbol{w})$ formed by four independent variables $w_0$ and $\boldsymbol{w}=(w^{1},w^{2},w^{3})$ is introduced, where the temporal component $w_0$ plays the role of an energy parameter.

In this paper, starting from the relation $y_{\mu}y^{\mu}=\tau^{2}$ we try to formally develop a duality theory of the usual quantum mechanics, which is different from the momentum-space representation of the usual quantum mechanics. The usual quantum mechanics is based on the relation $p^{\mu}p_{\mu}=m^{2}$ and describes "particle" states or mass eigenstates, while the duality theory is based on the relation $y_{\mu}y^{\mu}=\tau^{2}$ and describes "event" states or proper-time eigenstates. The usual quantum mechanics and its duality theory are regarded as two complementary aspects of a whole theory. The natural units of measurement ($\hbar=c=1$) are used and the space-time metric tensor is taken as $g_{\mu\nu}=\mathrm{diag}(1,-1,-1,-1)$ ($\mu,\nu=0,1,2,3$). Repeated indices denote summation according to the Einstein rule.

## 2. DUALITY THEORY OF THE USUAL MECHANICS

In order to introduce the duality theory of the usual quantum mechanics, let us firstly summarize the characteristics of the traditional mechanics as follows:

( Ⅰ ) The general formulation of the law governing the motion of mechanical systems is the principle of least action, where the Lagrangian is integrated over time to get the action (say, $A$). Let $L = L(Q,\dot{Q},t)$ be the Lagrangian of the system concerned, where $Q = (Q_1, Q_2, ...Q_n)$ and $\dot{Q} \equiv \mathrm{d}Q/\mathrm{d}t$ denote the generalized coordinates and generalized velocities, respectively. Let the system occupy, at the instants $t_1$ and $t_2$, positions denoted by $Q^{(1)}$ and $Q^{(2)}$, then the system moves between these positions in such a way that the integral (i.e. the action) $A = \int_{t_1}^{t_2} L(Q,\dot{Q},t)\mathrm{d}t$ takes the least possible value [8-9].

(Ⅱ) A position four-vector $x^{\mu} = (t, \boldsymbol{x})$ corresponds to four independent variables (where the time $t$ plays the role of a parameter and $t \in (-\infty, +\infty)$), while only three components of a 4-momentum $p^{\mu} = (E, \boldsymbol{p})$ are independent because of the mass-shell relation $p^{\mu} p_{\mu} = m^2$ ($m$ is a mass).

(Ⅲ) The Poincaré group has ten generators, which satisfy the Lie algebra of the group ($\mu, \nu, \lambda, \rho, \sigma = 0,1,2,3$)

$$\begin{cases} [\hat{p}_{\mu}, \hat{p}_{\nu}] = 0 \\ [\hat{J}_{\mu\nu}, \hat{p}_{\lambda}] = \mathrm{i}(g_{\lambda\nu}\hat{p}_{\mu} - g_{\lambda\mu}\hat{p}_{\nu}) \\ [\hat{J}_{\mu\nu}, \hat{J}_{\rho\sigma}] = -\mathrm{i}(g_{\mu\rho}\hat{J}_{\nu\sigma} + g_{\nu\sigma}\hat{J}_{\mu\rho} - g_{\nu\rho}\hat{J}_{\mu\sigma} - g_{\mu\sigma}\hat{J}_{\nu\rho}) \end{cases} \quad (1)$$

where $\hat{J}_{\mu\nu} = \hat{L}_{\mu\nu} + S_{\mu\nu}$ denotes the 4-tensor operator of total angular momentum, $S_{\mu\nu}$ represents the spin tensor, $\hat{L}_{\mu\nu} = x_{\mu}\hat{p}_{\nu} - x_{\nu}\hat{p}_{\mu}$ is the 4-tensor operator of orbital angular momentum and $\hat{p}_\mu = \mathrm{i}\partial/\partial x^\mu = \mathrm{i}\partial_\mu$ the 4-momentum operator. In the present case the 4-coordinate $x^\mu$ is not a generator of the Poincaré group.

(Ⅳ) Based on the traditional analytical mechanics, the quantum mechanics equation (or quantum field equation) can be given by the Schrödinger equation

$$\mathrm{i}\frac{\partial}{\partial t}\psi(t,\boldsymbol{x}) = \hat{H}\psi(t,\boldsymbol{x}) \qquad (2)$$

where $\hat{H}$ represents the Hamiltonian. Obviously, the usual quantum-mechanical equation (2) describes the time evolution of a "particle" state or a mass eigenstate that lies on a mass shell.

Here, it is very important for us to clarify a traditional misconception [4-5]. Let $\phi(x)$ stand for a field quantity and $m$ for its rest mass, using the commutation relation $[x^\mu, \hat{p}^\nu] = -\mathrm{i}g^{\mu\nu}$ one has $[\hat{p}^\nu \hat{p}_\nu, x^\mu] = 2\mathrm{i}\hat{p}^\mu$. Historically someone have obtained $[m^2, x^\mu] = 2\mathrm{i}\hat{p}^\mu$ from $\hat{p}^\mu \hat{p}_\mu \phi = m^2\phi$ and $[\hat{p}^\nu \hat{p}_\nu, x^\mu] = 2\mathrm{i}\hat{p}^\mu$, and arrived at a conclusion that a mass eigenstate does not correspond to a position eigenstate, which conflicts with the fact that a particle with a definite mass can be localized in space-time. Therefore they have concluded that the commutation relation $[x^\mu, \hat{p}^\nu] = -\mathrm{i}g^{\mu\nu}$ is not valid. However the objection against $[x^\mu, \hat{p}^\nu] = -\mathrm{i}g^{\mu\nu}$ is reasonless, because it is a purely mathematical truth rather than a physical assumption, and is always valid (it is $[x^\mu, \mathrm{i}\partial^\nu]\phi = -\mathrm{i}g^{\mu\nu}\phi$ by definition). In fact, only via $\hat{p}^\nu \hat{p}_\nu (x^\mu \phi) = m^2 (x^\mu \phi) (= x^\mu m^2 \phi = x^\mu \hat{p}^\nu \hat{p}_\nu \phi)$ can one obtain the erroneous result $[m^2, x^\mu] = 2\mathrm{i}\hat{p}^\mu$ from $\hat{p}^\mu \hat{p}_\mu \phi = m^2\phi$ and $[\hat{p}^\nu \hat{p}_\nu, x^\mu] = 2\mathrm{i}\hat{p}^\mu$ (it is $[\hat{p}^\nu \hat{p}_\nu, x^\mu]\phi = 2\mathrm{i}\hat{p}^\mu\phi$ by definition), but the equation $\hat{p}^\nu \hat{p}_\nu (x^\mu \phi) = m^2 (x^\mu \phi)$ is erroneous. Note that $\hat{p}^\mu \hat{p}_\mu \phi = m^2\phi$ is a physical equation rather than a mathematical identity $\hat{p}^\mu \hat{p}_\mu \equiv m^2$, one cannot obtain $\hat{p}^\nu \hat{p}_\nu (x^\mu \phi) = m^2 (x^\mu \phi)$ from $\hat{p}^\mu \hat{p}_\mu \phi = m^2\phi$, and the former contains an operation of $\partial^{\nu}$ with respect to $x^{\mu}$ such that one has $\hat{p}^{\nu}\hat{p}_{\nu}(x^{\mu}\phi) = 2\hat{p}^{\mu}\phi + x^{\mu}m^{2}\phi \neq x^{\mu}(m^{2}\phi)$. On the other hand, the mass $m$ is a c-number such that $[m^{2}, x^{\mu}] = 0 \neq 2\mathrm{i}\hat{p}^{\mu}$.

Now, a duality theory of the traditional mechanics could be constructed by replacing the above features (Ⅰ), (Ⅱ), (Ⅲ) and (Ⅳ) with the following features (ⅰ), (ⅱ), (ⅲ) and (ⅳ), respectively.

(ⅰ) The general formulation of the law governing the motion of mechanical systems is also the principle of least action, but it is the generalized Lagrangian $\overline{L} = \overline{L}(P, P', w_0)$ that is integrated over the energy coordinate $w_0$ to get the action $A = \int_{w_{01}}^{w_{02}} \overline{L}(P, P', w_0)\mathrm{d}w_0$, where $P = (P_1, P_2, ...P_n)$ is the generalized momentum and $P' = \mathrm{d}P/\mathrm{d}w_0$. The energy coordinate $w_0 \in (-\infty, +\infty)$ has the dimension of [1/length], and is also called *energy parameter*, of which we shall give a further discussion later.

(ⅱ) A 4D momentum coordinate formed by four independent variables $w_0$ and $\boldsymbol{w} = (w^1, w^2, w^3)$ is introduced. This four-vector has the dimension of [1/length], and the space spanned by $w^{\mu} = (w_0, \boldsymbol{w})$ is called the 4D momentum-coordinate space. On the other hand, only three components of a 4D space-time interval $y^{\mu} = (y^0, \boldsymbol{y})$ are independent because of $y_{\mu}y^{\mu} = \tau^2$, where $\tau$ is the proper-time interval. Here the time interval is treated as an operator on an equal footing with the space interval. Even though a coordinate plays the role of a parameter, the relevant coordinate interval can play the role of a dynamical variable. After all, time interval is an observable.

(ⅲ) If a physical quantity $M$ satisfies $\partial M/\partial w_0 = 0$, $M$ is called a conserved quantity with respect to $w_0$ (i.e., $M$ has a value constant in $w_0$). Using such a concept of conserved quantity and taking the 4D momentum-coordinate space $w^{\mu} = (w_0, \boldsymbol{w})$ as the representation space, the Poincaré group has ten generators, which satisfy the Lie algebra of the group ( $\mu,\nu,\lambda,\rho,\sigma = 0,1,2,3$ )

$$\begin{cases} [\hat{y}_\mu, \hat{y}_\nu] = 0 \\ [\hat{J}_{\mu\nu}, \hat{y}_\lambda] = \mathrm{i}(g_{\lambda\nu}\hat{y}_\mu - g_{\lambda\mu}\hat{y}_\nu) \\ [\hat{J}_{\mu\nu}, \hat{J}_{\rho\sigma}] = -\mathrm{i}(g_{\mu\rho}\hat{J}_{\nu\sigma} + g_{\nu\sigma}\hat{J}_{\mu\rho} - g_{\nu\rho}\hat{J}_{\mu\sigma} - g_{\mu\sigma}\hat{J}_{\nu\rho}) \end{cases} \tag{3}$$

where $\hat{J}_{\mu\nu} = \hat{L}_{\mu\nu} + S_{\mu\nu}$ denotes the 4-tensor operator of total angular momentum, $S_{\mu\nu}$ represents the spin tensor, $\hat{L}_{\mu\nu} = \hat{y}_\mu w_\nu - \hat{y}_\nu w_\mu$ is the 4-tensor operator of orbital angular momentum and $\hat{y}^\mu = -\mathrm{i}\,\partial/\partial w_\mu = -\mathrm{i}\partial_w^\mu$ is the space-time interval operator. In the present case the 4D momentum-coordinate $w^\mu$ is not a generator of the Poincaré group.

(iv) As a duality concept of the Hamiltonian in the usual mechanics, a *time function* (operator) $\hat{T}$ with the dimension of [length] is introduced. In general, one has $\hat{T} = \hat{T}(Q, P, w_0)$, where $Q$ and $P$ are the generalized coordinates and momenta, respectively. Analogous to the Schrödinger equation (2) which describes the time evolution of states $\psi(t, \boldsymbol{x})$, the energy translation of states $\phi(w_0, \boldsymbol{w})$ is described as:

$$-\mathrm{i}\frac{\partial}{\partial w_0}\phi(w_0, \boldsymbol{w}) = \hat{T}\phi(w_0, \boldsymbol{w}) \tag{4}$$

Choosing a right generalized Lagrangian $\bar{L} = \bar{L}(P, P', w_0)$ and applying the principle of least action in $A = \int_{w_{01}}^{w_{02}} \bar{L}(P, P', w_0)\mathrm{d}w_0$, one can obtain Eq. (4). Similarly, let $A[\boldsymbol{w}, \boldsymbol{y}]$ denote an action, one can give the path integral representation of energy translation amplitudes

$$\langle \boldsymbol{w}, w_0 | \boldsymbol{w}', w_0' \rangle = \int \frac{\mathrm{D}\boldsymbol{w}}{2\pi} \int \mathrm{D}\boldsymbol{y} \exp(\mathrm{i}A[\boldsymbol{w}, \boldsymbol{y}]) \tag{5}$$

The mechanics based on the above four features ( i ), ( ii ), (iii) and (iv) is then constructed and referred to the duality theory of the usual mechanics. In contrast to Eq. (2) describing the time evolution of a "particle" state (or a mass eigenstate), Eq. (4), as the duality counterpart of Eq. (2), describes that a proper-time eigenstate varies with the energy parameter.

Just as the relativistic dispersion relation $p^{\mu}p_{\mu}=m^{2}$ is called the mass-shell relation, we call $y_{\mu}y^{\mu}=\tau^{2}$ the proper-time-shell relation. Consider that the proper-time-shell relation is satisfied by the 4D space-time intervals between events, the proper-time eigenstate $\phi(w_{0},\boldsymbol{w})$ is also called an event eigenstate. Historically, some attempts of describing event eigenstates can be found in Ref. [4-5, 10], but they are based on the usual mechanics, and different from the duality theory given here.

## 3. ENERGY PARAMETER AND TIME FUNCTION

Now let us discuss the physical meanings of energy parameter and time function. As we know, when one puts a physical system in a potential field with the potential $V_0 \in(-\infty,+\infty)$ and $\partial V_{0}/\partial x^{\mu}=0$ ($\mu=0,1,2,3$), it is equivalent to choosing a new zero-energy reference point for the system. Seeing that all the physical laws are independent of a zero-energy reference point, one can introduce an energy coordinate or energy parameter $w_{0}=E+V\in(-\infty,+\infty)$, where $E$ is the usual energy. *Via* the potential $V_0$ a new degree of freedom is introduced, such that $w_0$ and $\boldsymbol{w}=(w^{1},w^{2},w^{3})$ form four independent variables, they do not satisfy the mass-shell relation. Moreover, assume that in a general inertial system, one has a 4D potential $V^{\mu}=(V_{0},\boldsymbol{V})$ with $\partial V^{\mu}/\partial x^{\nu}=0$ ($\mu,\nu=0,1,2,3$), such that one can choose the 4D momentum coordinate as

$$w^{\mu}=p^{\mu}+V^{\mu} \qquad (6)$$

where $p^{\mu}=(E,\boldsymbol{p})$ is the ordinary 4D momentum satisfying the mass-shell relation

$p^{\mu}p_{\mu}=m^2$. In the presence of the 4D potential $V^{\mu}=(V_0,\boldsymbol{V})$, the 4D momentum coordinate $w^{\mu}$ does not satisfy a fixed mass-shell relation, which due to the fact that, in a proper-time eigenstate, the mass would be indeterminate [6-7]. As we known, in the usual mechanics, a negative time represents the past while a positive time describes the future. Similarly, a negative energy-parameter represents the energy below a zero-energy reference point while a positive energy-parameter represents the energy above the zero-energy reference point.

For simplicity let us only consider the flat space-time. In the Minkowski space-time, a 4D line element d$s$ satisfies

$$(\mathrm{d}t)^2=(\mathrm{d}\boldsymbol{x})^2+(\mathrm{d}s)^2 \qquad (7)$$

Without loss of generality, let $\mathrm{d}t=\sqrt{(\mathrm{d}\boldsymbol{x})^2+(\mathrm{d}s)^2}$. For convenience, let us replace the differential elements $\mathrm{d}x^{\mu}$ and d$s$ with the corresponding intervals $y^{\mu}=(y^0,\boldsymbol{y})$ and $\tau$, respectively, then $y^0=\sqrt{(\boldsymbol{y})^2+(\tau)^2}$. Furthermore, assuming that the interval $y^{\mu}=(y^0,\boldsymbol{y})$ is time-like, in the nonrelativistic limit, one has

$$y^0=\sqrt{(\tau)^2+(\boldsymbol{y})^2}\approx\tau+\frac{(\boldsymbol{y})^2}{2\tau} \qquad (8)$$

Then, as a duality counterpart of kinetic energy in the usual mechanics, one can assume that, at the moment, the time function is

$$T=(\boldsymbol{y})^2/2\tau \qquad (9)$$

In order to obtain Eq. (4), we replace the 4D space-time interval $y^{\mu}$ by the corresponding operator $\hat{y}^{\mu}=-\mathrm{i}\partial/\partial w_{\mu}=-\mathrm{i}\partial_w^{\mu}$, where $\hat{\boldsymbol{y}}=\mathrm{i}\nabla_w=\mathrm{i}(\partial/\partial w^1,\partial/\partial w^2,\partial/\partial w^3)$ stands for the space-interval operator, and $\hat{T}=(\hat{\boldsymbol{y}})^2/2\tau$ for the time function operator. Then Eq. (4) becomes:

$$-\mathrm{i}\frac{\partial}{\partial w_0}\phi(w_0,\boldsymbol{w})=\frac{(\hat{\boldsymbol{y}})^2}{2\tau}\phi(w_0,\boldsymbol{w}) \qquad (10)$$

In the following we denote $\psi(x)=\psi(t,\boldsymbol{x})$ and $\phi(w)=\phi(w_0,\boldsymbol{w})$. In the relativistic case, in terms of the Dirac matrices $\gamma^\mu=(\beta,\boldsymbol{\gamma})$ and $\boldsymbol{\alpha}=\beta\boldsymbol{\gamma}$, for example, one has $(y^0)^2=(\boldsymbol{\alpha}\cdot\boldsymbol{y}+\beta\tau)^2$. Then the time function can be taken as $T=\boldsymbol{\alpha}\cdot\boldsymbol{y}+\beta\tau$ and the corresponding operator is $\hat{T}=\boldsymbol{\alpha}\cdot\hat{\boldsymbol{y}}+\beta\tau$, Eq. (4) becomes

$$-\mathrm{i}\frac{\partial}{\partial w_0}\phi(w)=(\boldsymbol{\alpha}\cdot\hat{\boldsymbol{y}}+\beta\tau)\phi(w) \qquad (11)$$

Or in the Lorentz-covariant form

$$(\mathrm{i}\gamma_\mu\partial_w^\mu-\tau)\phi(w)=0 \qquad (12)$$

Eq. (11) or (12) is easily obtained through applying the principle of least action in $A=\int \mathcal{L}\mathrm{d}^4w$, where

$$\mathcal{L}=\bar{\phi}(w)(\mathrm{i}\gamma_\mu\partial_w^\mu-\tau)\phi(w) \qquad (13)$$

is referred to as generalized Lagrangian density (with the dimension of $[\text{length}]^4$ instead of $[1/\text{length}]^4$), and $\bar{\phi}(w)=\phi^+(w)\gamma^0$ and $\phi^+(w)$ is the Hermitian conjugate of $\phi(w)$.

Analogously, in the simplest case, using $y_\mu y^\mu=\tau^2$ and $\hat{y}^\mu=-\mathrm{i}\partial_w^\mu$, one can directly introduce an equation

$$(\partial_w^\mu\partial_{w\mu}+\tau^2)\phi(w)=0 \qquad (14)$$

Applying the principle of least action in $A=\int \mathcal{L}\mathrm{d}^4w$, Eq. (14) is easily obtained, where

$$\mathcal{L}=\frac{1}{2}[\partial_w^\mu\phi(w)][\partial_{w\mu}\phi(w)]-\frac{1}{2}\tau^2\phi^2(w) \qquad 15)$$

is the generalized Lagrangian density (with the dimension of $[\text{length}]^4$).

Moreover, we call the duality counterpart of potential energy the “time potential”.

## 4. VARIABLE-MASS PARAMETER FORMULATION

As well known [9-11], in term of a continuous Poincaré-invariant parameter $s$ (i.e., the proper time), the covariant Lagrange equations of a system of $N$ particles are ($i=1,2,3,...N$, $\mu=0,1,2,3$)

$$\frac{\mathrm{d}}{\mathrm{d}s}\left(\frac{\partial L}{\partial \dot{q}_i^\mu}\right)-\frac{\partial L}{\partial q_i^\mu}=0 \qquad (16)$$

where $q_i^\mu$ are the 4D generalized coordinates, $\dot{q}_i^\mu=\partial q_i^\mu/\partial s$ the 4D generalized velocities, and $L$ the Lagrangian. Correspondingly, the covariant Hamilton equations are

$$\frac{\mathrm{d}q_i^\mu}{\mathrm{d}s}=\frac{\partial H}{\partial p_{i\mu}},\quad \frac{\mathrm{d}p_i^\mu}{\mathrm{d}s}=-\frac{\partial H}{\partial q_{i\mu}} \qquad (17)$$

where $p_i^\mu$ are the 4D generalized momentums and $H$ the Hamiltonian.

Likewise, using Eq. (6) one can define a Poincaré-invariant parameter $m_V$

$$m_V\equiv\sqrt{w^\mu w_\mu} \qquad (18)$$

it is referred to as variable-mass parameter. In terms of the variable-mass parameter, in our duality theory, the dual counterparts of Eqs. (16) and (17) can be written as, respectively

$$\frac{\mathrm{d}}{\mathrm{d}m_V}\left(\frac{\partial \bar{L}}{\partial w_i'^\mu}\right)-\frac{\partial \bar{L}}{\partial w_i^\mu}=0 \qquad (19)$$

$$\frac{\mathrm{d}w_i^\mu}{\mathrm{d}m_V}=\frac{\partial T}{\partial y_{i\mu}},\quad \frac{\mathrm{d}y_i^\mu}{\mathrm{d}m_V}=-\frac{\partial T}{\partial w_{i\mu}} \qquad (20)$$

where $w_i^\mu$ are the generalized momentum-coordinates, $y_i^\mu$ the generalized coordinate-intervals, $w_i'^\mu=\partial w_i^\mu/\partial m_V$, $T$ is the time function, and $\bar{L}$ the generalized Lagrangian with the dimension of [length]. Here note that, from the usual mechanics to our duality theory, one must make some transformations such as ($\hbar=c=1$)

$$\begin{cases} \boldsymbol{x} = \boldsymbol{v}t \Leftrightarrow \boldsymbol{w} = \boldsymbol{v}w_0/c^2 = \boldsymbol{v}w_0 \\ \mathrm{d}\boldsymbol{x}/\mathrm{d}t = \boldsymbol{v} \Leftrightarrow \mathrm{d}\boldsymbol{w}/\mathrm{d}w_0 = \boldsymbol{v}/c^2 = \boldsymbol{v} \end{cases} \tag{21}$$

$$\begin{cases} E = \sqrt{\boldsymbol{p}^2 + m^2} \Leftrightarrow T = \sqrt{\boldsymbol{y}^2 + \tau^2} \\ \mathrm{d}E/\mathrm{d}\boldsymbol{p} = \boldsymbol{p}/E = \boldsymbol{v}/c^2 = \boldsymbol{v} \Leftrightarrow \mathrm{d}T/\mathrm{d}\boldsymbol{y} = \boldsymbol{y}/T = \boldsymbol{v} \end{cases} \tag{22}$$

In fact, the concept of variable mass or indefinite mass can be found in previous literatures [12-19], though it was introduced for different purposes and has different physical interpretations in different literatures.

Though the arguments given in this section are purely mathematical, in **Appendix A** we will give a heuristic justification for this section.

## 5. SPACE-TIME INTERVAL QUANTIZATION

Using the usual quantum theory, one can find the phenomena of energy and momentum quantization. Now, in our duality theory, we will discuss the quantization of space-time interval. As an example, we take Eqs. (14) and (15) as our starting point. As mentioned before, if a physical quantity $M$ satisfies $\partial M/\partial w_0 = 0$, $M$ is defined as a conserved quantity with respect to $w_0$. Using such a conservation concept one can show that, under an infinitesimal transformation $w_\mu \to w'_\mu = w_\mu + k_\mu$ ($k_\mu$ is an infinitesimal constant), the generalized Lagrangian density $\pounds$ given by Eq. (15) is invariant. Correspondingly, there are four generalized conserved quantities (say, $\hat{Y}^\mu = (\hat{T}, \hat{\boldsymbol{Y}})$), that is $\partial \hat{Y}^\mu/\partial w_0 = 0$ ($\mu = 0,1,2,3$), where

$$\hat{T} = \int [\pi(w)\phi'(w) - \pounds]\mathrm{d}^3 w \tag{23}$$

$$\hat{\boldsymbol{Y}} = -\int [\pi(w)\nabla_w \phi(w)]\mathrm{d}^3 w \tag{24}$$

where $\phi'(w) = \partial\phi(w)/\partial w_0$, $\pi(w)$ is the canonically conjugate field of $\phi(w)$, i.e.,

$$\pi(w) = \frac{\partial \pounds}{\partial \phi'(w)} = \phi'(w) \tag{25}$$

Let $\int \mathrm{d}^3 w \equiv V_w$, the general solution of Eq. (14) can be expanded as

$$\phi(w) = \sum_{y} \sqrt{\frac{1}{2y^0 V_w}} [a(y)\exp(-\mathrm{i}w \cdot y) + a^+(y)\exp(\mathrm{i}w \cdot y)] \qquad (26)$$

The expansion coefficients $a(y)$ and $a^+(y)$ (Hermitian conjugate of $a(y)$) are called the annihilation and the creation operators of events.

Now we postulate that these field operators satisfy the equal-$w_0$ commutation rules

$$\begin{cases} [\phi(w_0, \boldsymbol{w}), \pi(w_0, \boldsymbol{w}')] = \mathrm{i}\delta^3(\boldsymbol{w} - \boldsymbol{w}') \\ [\phi(w_0, \boldsymbol{w}), \phi(w_0, \boldsymbol{w}')] = [\pi(w_0, \boldsymbol{w}), \pi(w_0, \boldsymbol{w}')] = 0 \end{cases} \qquad (27)$$

Using Eqs. (25)-(27), one has

$$[a(\boldsymbol{y}), a^+(\boldsymbol{y}')] = \delta_{\boldsymbol{y}\boldsymbol{y}'} \qquad (28)$$

with the others vanishing. With Eqs. (23-28), one gets

$$\hat{T} = \sum_{y} y_0 [a^+(y)a(y) + \frac{1}{2}] \qquad (29)$$

$$\hat{\boldsymbol{Y}} = \sum_{y} \boldsymbol{y}[a^+(y)a(y)] \qquad (30)$$

Eqs. (29)-(30) describe the quantization of the space-time intervals between the events. It is interesting to note that there is a zero-point time-interval presented in Eq. (29).

## 6. DISCUSSIONS AND PROSPECTS

Eq. (29) shows that there is a zero-point time-interval. This result is valid for general space-time for the reason that the flat space-time is a special case of the general space-time. Here we only give a mathematical model and formal theory. In our next work, we will try to give a concrete physical model. Specifically, by the aid of the concept of zero-point time-interval, we will try to give an alternative interpretation for the spontaneous emission and the lifetime of excited state. As we know, the presence of zero-point energy has an effect on the lifetime of excited state and induces a spontaneous emission. On the other hand, owing to the time-energy uncertainty relation, a zero-point fluctuation of energy must be accompanied with a zero-point fluctuation of some kind of time scale, we will try to identify this time scale with the zero-point time-interval.

**Acknowledgements**. We would like to acknowledge the financial support of the Doctoral Program Foundation of Institution of Higher Education of China (Grant No. 20050614022).

## APPENDIX A

In order to show some possible physical reality, here we shall give a heuristic justification for Section 4. However, a rigorous and concrete physical model remains to be given in future. Let $A(p,q,t)$ be an action, where $q=(q_1,q_2...,q_n)$ and $p=(p_1,p_2...,p_n)$ are the generalized coordinates and momenta, respectively. By the Jacobi-Hamilton equation

$$\frac{\partial A}{\partial t}+H=0 \qquad \text{(A1)}$$

one can define the Hamiltonian *H*. The Hamilton equations are

$$\begin{cases} \frac{\partial p_i}{\partial t}=-\frac{\partial H}{\partial q_i}=\frac{\partial^2 A}{\partial q_i \partial t}=\frac{\partial}{\partial t}(\frac{\partial A}{\partial q_i}) \\ \frac{\partial q_i}{\partial t}=\frac{\partial H}{\partial p_i}=-\frac{\partial^2 A}{\partial p_i \partial t}=-\frac{\partial}{\partial t}(\frac{\partial A}{\partial p_i}) \end{cases} \quad (i=1,2,...n) \qquad \text{(A2)}$$

Then

$$p_i=\frac{\partial A}{\partial q_i}+C_{1i},\ \ q_i=-(\frac{\partial A}{\partial p_i}+C_{2i}) \quad (i=1,2,...n) \qquad \text{(A3)}$$

where $C_{1i}$ and $C_{2i}$ ( $i=1,2,...n$ ) are constants, for simplicity, let $C_{1i}=C_{2i}=0$ ( $i=1,2,...n$ ), then

$$q_i=-\frac{\partial A}{\partial p_i},\quad p_i=\frac{\partial A}{\partial q_i} \quad (i=1,2,...n) \qquad \text{(A4)}$$

Now defining the time function (as a heuristic argument, we directly take the Hamiltonian *H* as the energy parameter)

$$T\equiv-\frac{\partial A}{\partial H} \qquad \text{(A5)}$$

Using Eqs. (A4) and (A5), one has

$$\begin{cases} \dfrac{\partial T}{\partial q_i} = -\dfrac{\partial}{\partial H}(\dfrac{\partial A}{\partial q_i}) = -\dfrac{\partial p_i}{\partial H} \\ \dfrac{\partial T}{\partial p_i} = -\dfrac{\partial}{\partial H}(\dfrac{\partial A}{\partial p_i}) = \dfrac{\partial q_i}{\partial H} \end{cases} \quad (i = 1, 2, ...n) \qquad \text{(A6)}$$

That is

$$\begin{cases} \dfrac{\partial p_i}{\partial H} = -\dfrac{\partial T}{\partial q_i} \\ \dfrac{\partial q_i}{\partial H} = \dfrac{\partial T}{\partial p_i} \end{cases} \qquad \text{(A7)}$$

Obviously, Eq. (A7) is to Eq. (A2) as Eq. (20) is to Eq. (17).